# A Diamond Nanowire Single Photon Antenna

Tom M. Babinec[1], Birgit J. M. Hausmann[1,2], Mughees Khan[1], Yinan Zhang[1], Jero Maze[3], Phil R. Hemmer[4], Marko Lončar[1*]

1. School of Engineering and Applied Science, Harvard University, Cambridge, MA 02138, USA.

2. Department of Physics, Technische Universität München, D-85748 Garching, Germany

3. Department of Physics, Harvard University, Cambridge, MA 02138, USA.

4. Department of Electrical and Computer Engineering, Texas A&M University, College Station, TX 77843, USA.

**The development of a robust light source that emits one photon at a time is an outstanding challenge in quantum science and technology. Here, at the transition from many to single photon optical communication systems, fully quantum mechanical effects may be utilized to achieve new capabilities, most notably perfectly secure communication via quantum cryptography[1]. Practical implementations place stringent requirements on the device properties, including stable photon generation, room temperature operation, and efficient extraction of many photons. Single photon light emitting devices based on fluorescent dye molecules[2], quantum dots[3], and carbon nanotube[4] material systems have all been explored, but none have simultaneously demonstrated all criteria. Here, we**

[*] To whom correspondence should be addressed: loncar@seas.harvard.edu

**describe the design, fabrication, and characterization of a bright source of single photons consisting of an individual Nitrogen-vacancy color center (NV center) in a diamond nanowire operating in ambient conditions. The nanowire plays a positive role in increasing the number of single photons collected from the NV center by an order of magnitude over devices based on bulk diamond crystals, and allows operation at an order of magnitude lower power levels. This result enables a new class of nanostructured diamond devices for room temperature photonic and quantum information processing applications, and will also impact fields as diverse as biological and chemical sensing, opto-mechanics, and scanning-probe microscopy.**

The variegated defect centers in diamond, including Si[5], Nickel[6], Cr[7], and Nitrogen[8] complexes, have recently emerged as a solid-state platform for single-photon source applications. Of these, the negatively charged Nitrogen-vacancy color center, which is composed of an adjacent substitutional Nitrogen atom and diamond lattice vacancy (Fig. 1a), has generated much interest because of its potential for combining photonic and spin qubits in an on-chip device architecture even at room temperature[9]. The NV center has its own electronic structure that may be optically addressed, and early studies in a bulk diamond crystal have already demonstrated optical read-out of the electronic spin state[10], coupling of electron and nuclear spins in a quantum register[11], multipartite entanglement[12], and nanoscale magnetic field sensing[13,14]. For devices based on an NV center in a bulk diamond crystal, the majority of emitted photons are lost to the substrate via total internal reflection at the diamond-air interface. A device platform that can be used to efficiently probe an individual NV center therefore represents a major opportunity for wider applications of diamond in quantum information processing (QIP) applications.

The performance of diamond-based QIP systems may be improved by embedding color centers in nanophotonic structures. The main engineering challenge is to position an NV center at a location of intense optical fields. One approach considered recently is to evanescently couple a separate optical cavity or a waveguide to a proximal NV center [15-19]. While this scenario utilizes mature fabrication schemes, light-matter interactions may be limited by reduced overlap of the NV center with the field maximum inside the device. Another approach is to realize optical structures directly in thin diamond films grown on foreign (low-index or sacrificial) substrates. At the present time only nano-crystalline diamond films are available and their optical properties are inferior to those of single crystal diamond[20]. For example, scattering of light off grain boundaries as well as background luminescence may render these nano-crystalline films incompatible for applications in quantum information processing. Finally, optical nanostructures may be sculpted from a bulk single crystal diamond[21-25]. Diamond nanofabrication presents a major challenge, though, and no techniques have yet been demonstrated that are compatible with requirements needed for realization of devices for quantum cryptography and QIP.

The approach taken in this Letter is to define large arrays of vertically oriented nanowires (Fig 1b) in a single crystal diamond substrate using a top-down nanofabrication technique. Three dimensional finite-difference time-domain (FDTD) calculations show that the performance of a single photon device based upon an NV center in a diamond nanowire antenna is dramatically improved relative to the bulk. For example, the fundamental nanowire mode ($HE_{11}$) is the dominant emission channel for an s-polarized dipole (polarized in the nanowire cross section) positioned on the nanowire axis (Fig. 1c)[26]. Light is emitted vertically from the top nanowire facet, which allows for efficient collection using an objective lens positioned above. This is true even in the case of a p-polarized dipole (polarized along the nanowire axis) that cannot emit into the fundamental mode, due to symmetry mismatch, which instead emits into vertically propagating radiation modes (Fig 1d). In this experiment the NV center dipole transition is polarized in the (111) plane of a (100) diamond nanowire[27,28], so that the average antenna effect on both s- and p- polarized contributions must be calculated. In addition, the overlap between the far-field profile of each component and the acceptance angle of our objective lens varies for different dipole wavelengths, which can be significant for the broad NV center emission. Finally, a slightly reduced lifetime is expected that is intermediate between the bulk diamond lifetime, where the lifetime is fast (~12ns) due to the large background refractive index, and the slow diamond nanoparticle lifetime (~25ns) of a dipole in air[25,26,29]. After averaging over the polarization and wavelength of these effects for the NV center, our simulations indicate that it is possible to extract ~30% of emitted photons from an NV center in a diamond nanowire. Overall, an order of magnitude improvement in the brightness of the nanowire-based single photon source is expected compared to the bulk.

The modest top-down nanofabrication requirements necessary to realize a diamond nanowire are an additional advantage of this platform[25]. First, a resist layer was deposited on a commercially available Type Ib diamond crystal, with (100) orientation. Next, electron-beam lithography was used to define circular masks with ~200nm diameters on the surface. Finally, a dry etch was used to transfer the pattern into the diamond crystal and realize straight wires of variable lengths ranging from 1-4 m with smooth sidewalls (Fig. 1e). Fabrication details are provided in the "Methods" section. Since the diamond used in this experiment possesses NV centers distributed throughout the crystal (result of the growth process), fabricated nanowires structures contain NV centers that are randomly positioned in the nanowire cross-section. The top-down process used to generate nanowires therefore mechanically isolates individual quantum emitters and minimizes background fluorescence.

The optical properties of an NV center in a diamond nanowire were studied using a laser scanning confocal microscope (Fig. 2a and described elsewhere[8]). A high-throughput technique was first used to scan large arrays of nanowires (Fig. 2b) and to identify the brightest single photon sources with the highest count rates (Fig. 2c). These devices could be studied in ambient conditions for long periods of time (days) due to the structural stability of the diamond nanowire and photo stability of the embedded NV center. Photoluminescence measurements of the emission spectrum from a typical nanowire device confirm the presence of embedded NV centers (Fig. 2d). The critical feature is a zero-phonon line at 637nm that is observed even at room temperature, as well as a broad spectrum of phonon assisted transitions present in the ~640-780nm wavelength.

Intensity autocorrelation measurements, performed using Hanbury Brown and Twiss method, confirm that one photon is emitted from the nanowire antenna at a time. In this experiment (Fig. 3a-c), collected light passes through a 50-50 beamsplitter and is detected with an avalanche photodiode (APD) at each output. Since a single photon cannot travel simultaneously through both arms of the interferometer, a delay time $\tau > 0$ is expected for a photon to be registered on both channels. The probability of measuring two photons delayed by a time $\tau$ is given by $g^{(2)}(\tau) = \langle I(t)\, I(t+\tau)\rangle / \langle I(t)\rangle^2$, which will ideally go to zero for $\tau = 0$ delay. For our nanowire device, strong photon antibunching ($g^{(2)}(0) < 1/2$) indicates that the coupling between one NV center and the nanowire antenna dominates all other background sources, including stray light, APD dark counts, and emission from an ensemble of NV centers in the substrate. At higher pump powers, the autocorrelation measurements exhibits bunching ($g^{(2)}(\tau) > 1$) at intermediate delay times due to optical cycling through a long-lived, non-radiative shelving state (Fig. 3b, c)[29]. In addition, the main features of the level crossing system that lead to the polarization mechanism of the $m_s = 0$ sublevel of the triplet ground state and the spin-dependent fluorescence rate remain unchanged after nanostructuring[30], as confirmed by standard ESR and Rabi measurements (data not shown).

The fluorescence lifetime of a color center in the diamond nanowire gives an upper bound on the number of single photons that may be collected. This is encoded in the temporal width of the autocorrelation data, and in particular the exponential decay of the antibunching dip has the form $e^{-(r+\Gamma)|\tau|}$ for low pump powers[8,29]. Here, r is the pump rate, $\Gamma = 1/\tau_{NW}$ is the NV center decay rate, and $\tau$ is the NV center lifetime. The overall decay rate $R = r + \Gamma$ was measured at different pump powers and observed to decrease linearly at low pump powers. The fitted decay rate presented in figure 3d gives a fluorescence lifetime of ~14.0ns. We tested total of six different nanowires and found the radiative lifetime to be in the range $\tau_{NW} \in (12.8\text{ns}, 16.5\text{ns})$, which is slightly longer the lifetime of an NV center in a bulk diamond[8] ($\tau_{NW}=11.8$ns), as expected.

The extent of the coupling between NV center and nanowire antenna determines the overall brightness of the single photon source. This can be obtained by measuring the number of single photon counts per second (cps) from the nanowire for different pump powers in a single photon L-L curve (Fig. 4). After a sharp rise at low pump powers (P < $P_{Sat}$), the number of collected photons per second saturates ($I_{Sat}$) at high powers due to the finite NV center emission rate. This can be fit to the form $I(P) = I_{sat} / (1 + P_{Sat}/P)$, where $P_{Sat}$ indicates how much optical power must be used to saturate the NV center response, and $I_{sat}$ indicates how many single photons may be collected from the device[8]. This data is presented for single NV centers in a bulk diamond crystal in figure 4a and in diamond nanowire devices in figure 4b. We found $I_{Sat}$ in the range 131 kcps to 205kcps, and $P_{Sat}$ in the range 21 W to 95 W for nanowire devices. At the same time, for bulk devices tested in this experiment $I_{Sat}$ was in the range 19 kcps to 23 kcps, for $P_{Sat}$ in the range 450 W to 1,530 W. Deviation from theoretical maximum count levels can be attributed to a variety of factors, including optical cycling through the metastable shelving state and losses in the optical system used in the experiment. Nevertheless, efficient in- and out-coupling of light from the diamond nanowire antenna provides an order of magnitude brighter single photon source operating at low power levels.

For the first time, we have utilized a top-down nanofabrication technique to enhance the optical properties of an individual color center in single crystal diamond. The fabrication technique maintains crucial properties of an NV center, and is compatible with requirements needed for realization of scalable quantum systems based on diamond. Further fundamental studies of the properties of an NV center in diamond nanostructures will facilitate their integration into more complex photonic devices, where more advanced functions such as increasing photon production rate via the Purcell effect will enable devices operating at even higher count levels and lower powers. Finally, by fabricating nanowires in ultra-pure diamond with implanted color centers, the simultaneous enhancement of spin and optical properties will be possible in a single device.

**Methods**

The single photon devices used in this experiment were prepared from a high-pressure high-temperature (HPHT) Type Ib diamond (Element 6 Corporation). The diamond was first cleaned in a boiling 1:1:1 Nitric, Perchloric, and Sulfuric acid bath to remove surface contamination. A 1:2 dilution of FOx 17 negative e-beam resist with methyl isobutyl ketone (MIBK) was then spun on the diamond to form the resist layer. Arrays of ~200 diameter circles were then patterned using an Elionix e-beam writing system at a dosage of ~6000 C/cm$^2$. 25% Tetra-methyl ammonium hydroxide (TMAH) was used to develop the resist and form the etch mask. The diamond crystal was then placed in a ICP RIE system and etched for 10 minutes with 30 sccm of Oxygen gas, 100W bias power at a 10mTorr chamber pressure. For the first two minutes 700W ICP power was applied, then three minutes of 600W ICP power, and finally five minutes at 1000W ICP power, resulting in vertical nanowire structures with ~2.0 m lengths. An HF wet etch was used to remove the mask from the top of the nanowires, and an additional acid bath treatment was performed prior to device testing.

# References

[1] Beveratos, A. *et al.* Single photon quantum cryptography. *Phys. Rev. Lett.* **89**, 187901 (2002).

[2] Lounis, B. & Moerner, W. E. Single photons on demand from a single molecule at room temperature. *Nature* **407**, 491 (2000).

[3] Englund, D. *et al.* Controlling the spontaneous emission rate of single quantum dots in a two-dimensional photonic crystal. *Phys. Rev. Lett.* **95**, 013904 (2005).

[4] Högele, A., Galland, C., Winger, M. & Imamoğlu, A. Photon antibunching in the photoluminescence spectra of a single carbon nanotube. *Phys. Rev. Lett.* **100**, 217401 (2008).

[5] Wang, C., Kurtsiefer, C., Weinfurter, H. & Burchard, B. Single photon emission from SiV centers in diamond produced by ion implantation. *J. Phys. B: At. Mol. Opt. Phys.* **39**, 37 (2006).

[6] Gaebel, T. *et al.* Stable single-photon source in the near infrared. *New Journal of Physics* **6**, 98 (2004).

[7] Aharonovich, I. *et al.* Two-level ultrabright single photon emission from diamond nanocrystals. *Nano Letters* **9**, 3191 (2009).

[8] Kurtsiefer, C., Mayer, S., Zarda, P. & Weinfurter, H. Stable solid-state source of single photons. *Phys. Rev. Lett.* **85**, 290 (2000).

[9] Prawer, S. & Greentree, A. D. Diamond for quantum computing. *Science* **320**, 1601 (2008).

10	Jelezko, F. *et al.* Observation of coherent oscillations in a single electron spin. *Phys. Rev. Lett.* **92**, 076401 (2004).

11	Gurudev Dutt, M. V. *et al.* Quantum register based on individual electronic and nuclear spin qubits in diamond. *Science* **316**, 1312 (2007).

12	Neumann, P. *et al.* Multipartite entanglement among single spins in diamond. *Science* **320**, 1326 (2008).

13	Maze, J. R. *et al.* Nanoscale magnetic sensing with an individual electronic spin in diamond. *Nature* **455**, 644 (2008).

14	Balasubramanian, G. *et al.* Nanoscale Imaging Magnetometry with Diamond Spins Under Ambient Conditions. *Nature* **455**, 648 (2008).

15	Park, Y.-S., Cook, A. K. & Wang, H. Cavity QED with diamond nanocrystals and silica microspheres. *Nano Lett.* **6**, 2075 (2006).

16	Larsson, M., Dinyari, K. N. & Wang, H. Composite optical microcavity of diamond nanopillar and silica microsphere. *Nano Lett.* **9**, 1447 (2009).

17	Fu, K.-M. C. *et al.* Coupling of nitrogen-vacancy centers in diamond to a GaP waveguide. *Appl. Phys. Lett.* **93**, 234107 (2008).

18	Barclary, P. E., Fu, K.-M. C., Santori, C. & Beausoleil, R. G. Chip-based microcavities coupled to NV centers in single crystal diamond. *arXiv:0908.2148*.

19	Kolosev, R. *et al.* Wave-particle duality of single surface plasmon polaritons. *Nature Physics* **5**, 470 (2009).

20    Wang, C. F. *et al.* Fabrication and characterization of two-dimensional photonic crystal microcavities in nanocrystalline diamond. *Appl. Phys. Lett.* **91**, 201112 (2007).

21    Lee, C. L. *et al.* Fabrication and characterization of diamond micro-optics. *Diamond and Related Materials* **15**, 725 (2006).

22    Wang, C. F., Hu, E. L., Yang, J. & Butler, J. E. Fabrication of Suspended Single Crystal Diamond Devices by Electrochemical Etch. *J. Vac. Sci. Technol. B* **25**, 730 (2007).

23    Fairchild, B. A. *et al.* Fabrication of ultrathin single-crystal diamond membranes. *Adv. Mater.* **20**, 4793 (2008).

24    Hisckocks, M. P. *et al.* Diamond waveguides fabricated by reactive ion etching. *Optics Express* **16**, 19512 (2008).

25    Hausmann, B. *et al.* Fabrication of Diamond Nanowires for Quantum Information Processing Applications. *arXiv:0908.0352*.

26    Friedler, I. *et al.* Solid-state single photon sources: the nanowire antenna. *Optics Express* **17**, 2095 (2009).

27    Epstein, R. J., Mendoza, F. M., Kato, Y. K. & Awschalom, D. D. Anisotropic interactions of a single spin and dark-spin spectroscopy in diamond. *Nature Physics* **1**, 94 (2005).

28    Gali, A., Fyta, M. & Kaxiras, E. *Ab initio* supercell calculations on nitrogen-vacancy center in diamond: Electronic structure and hyperfine tensors. *Phys. Rev. B* **77**, 155206 (2008).

29  Beveratos, A. *et al.* Nonclassical radiation from diamond nanocrystals. *Phys. Rev. A* **64**, 061802(R) (2001).

30  Manson, N. B., Harrison, J. P. & Sellars, M. J. Nitrogen-vacancy center in diamond: model of the electronic structure and associated dynamics. *Phys. Rev. B* **74**, 104303 (2006).

**Declaration of Financial Interests**

The authors declare no financial interests.

**Acknowledgments**

We would like to thank Florian Huber for providing Figure 1b. Helpful discussions with Qimin Quan, Emre Togan, Irfan Bulu, Mikhail Lukin, and Fedor Jelezko are acknowledged. We would also like to thank Sungkun Hong, Mike Grinolds, Patrick Maletinsky, and Amir Yacoby for confirmation of the ESR signal. Devices were fabricated in the Center for Nanoscale Systems (CNS) at Harvard. This work was supported in part by Harvard's Nanoscale Science and Engineering Center (NSEC), NSF NIRT grant (ECCS-0708905), and by the DARPA QuEST program. TB is funded by the NDSEG fellowship.

**Figure 1: Single photon source based on an NV center in diamond nanowire. a.** Cartoon of the Nitrogen Vacancy (NV) center, which consists of adjacent substitutional Nitrogen atom and lattice vacancy in the diamond lattice. Broadband single photon emission is due to a dipole transition that is polarized in the (111) crystal plane[28] (shaded triangle). **b.** Schematic representation of the device platform based on diamond crystal with randomly distributed NV centers (illustrated with light red dots). Photons emitted from NV center embedded within nanowire are directed towards the collection optics. **c.** Representative field profile of a radial component of E-field ($E_r$) in the case of a 2 m long, 200nm diameter diamond nanowire with an on-axis s-polarized dipole emitting at λ = 637nm (zero-phonon line wavelength of NV center), positioned at the nanowire center. The fundamental nanowire waveguide mode ($HE_{11}$) is the dominant decay channel for s-polarized dipole. Highly directional emission from the nanowire's top facet, within collection angle (θ ~ 72 ) of an objective lens with numerical aperture NA = 0.95 positioned above the nanowire, results in ~100% collection efficiency of emitted photons. **d.** Representative $E_r$ field profile for the same device but with a p-polarized dipole (vertical red arrow) positioned on the nanowire axis. Though in this case single photon emission does not couple to the waveguide mode, upward propagating radiation modes still allow for significant collection from an objective lens. **e.** Scanning electron microscopy (SEM) micrograph of a large array of fabricated diamond nanowire antennas.

**Figure 2: Confocal microscopy of an array of diamond nanowires containing NV centers. a.** The room temperature scanning confocal microscope used in this experiment consists of a 532nm CW laser used to excite the NV center, a color mirror to spectrally separate the red NV center, and a single mode fiber acting as a confocal pinhole to reject unfocused light. An air objective lens with NA=0.95 is used to focus the pump green light and collected emitted red photons. **b.** An SEM micrograph shows an array of typical ~200nm diameter nanowires analyzed in this experiment. **c.** Confocal microscope image of a square array of nanowire devices (5 m scale bar). Light blue and yellow spots correspond to nanowires with no embedded NV center and nanowires containing a weakly coupled NV center, respectively. Single photon sources with the best performance appear red in this image due to their high photon count rates. Inset shows blow-up of one of the nanowires, studied in this work. **d.** The photoluminescence spectrum of photons collected from a typical diamond nanowire shows the NV center zero-phonon line at ~637nm and phonon sideband from 640-780nm.

**Figure 3: Non-classical light emission from an NV center in a diamond nanowire. a-c.** The presence of a single quantum emitter in the diamond nanowire is revealed by the second-order autocorrelation function $g^{(2)}(\tau)$. The dramatic decrease in the number coincidence counts at zero time delay ($g^{(2)}(0)<1/2$) indicates that photons from the nanowire device are anti-bunched. Qualitatively different dynamics are observed at different excitation powers: (a) 11 W – below saturation, (b) 190 W – near (but above) saturation, and (c) 1.6mW – above saturation. At high pump powers, coupling to the metastable shelving state is significant and results in bunching shoulders ($g^{(2)}(\tau)>1$) at intermediate times. Red curves in (a) and (b) are fits using a three-level model of the $g^{(2)}(\tau)$ function[8]. **d.** The decay rate of the $g^{(2)}(\tau)$ spectrum measured for different excitation conditions gives the fluorescence lifetime in the limit of zero pump power[29]. This extrapolation yields a lifetime of $\tau = 14.0$ns for this NV-nanowire system, which is slightly larger than that of bulk material (11.8ns). This suggests slight suppression of emission in nanowire, consistent with our numerical model. Error bars represent $\pm 1\text{ns}^{-1}$ standard error in the decay rate fitting parameter.

**Figure 4: Comparison between diamond nanowire and bulk diamond crystal. a-b.** The number of single photons per second collected from a single NV center in an (a.) ultra-pure bulk diamond crystal and in a (b.) diamond nanowire (the same one as in Fig. 3) fabricated in diamond containing a high density of NV centers. Power was measured in front of the microscope objective. Black data shows raw count data from the device, hollow circles show the linear background data measured off the device, and colored circles shows the net NV center counts. The NV center L-L curves are fitted using the saturation model presented in the text. **c.** Distribution of count levels and saturation powers for the bulk diamond devices and diamond nanowire antennas tested in this experiment. The diamond nanowire antenna geometry allows roughly an order of magnitude more single photons to be extracted from an embedded NV center at an order of magnitude less pump power.

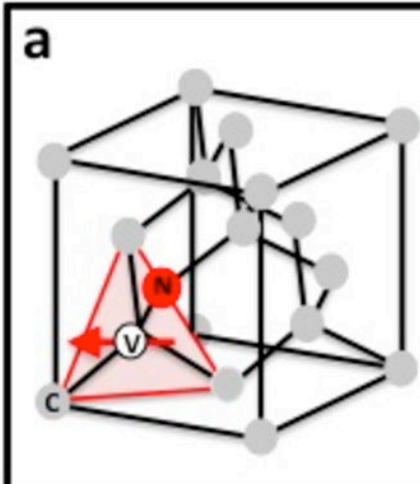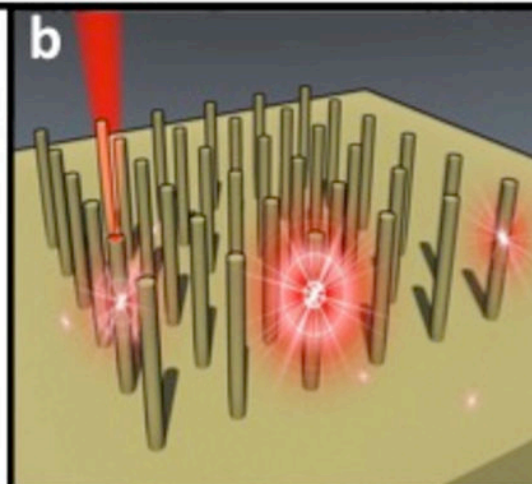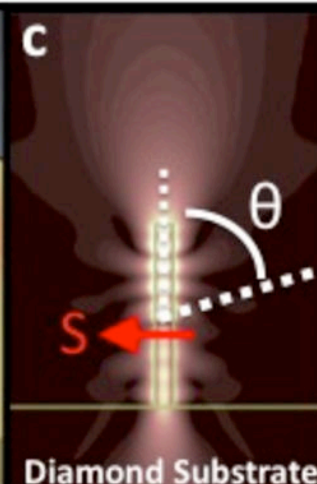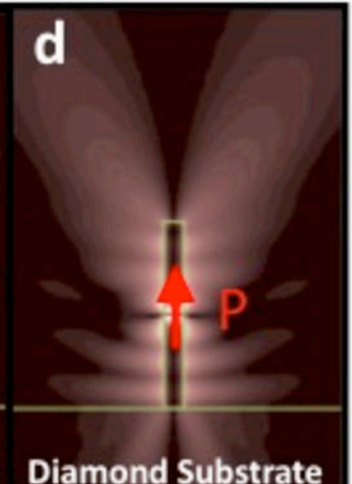
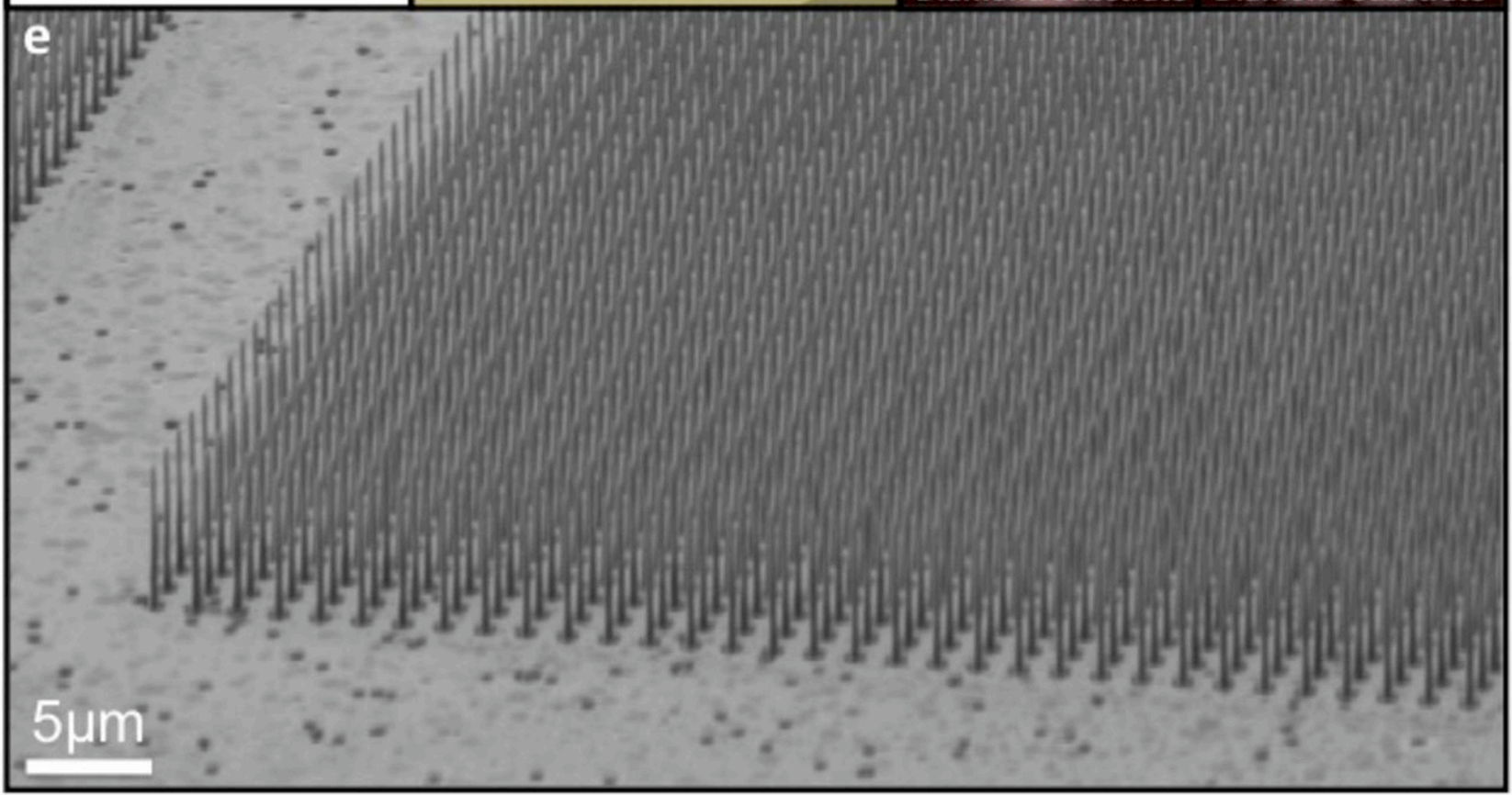

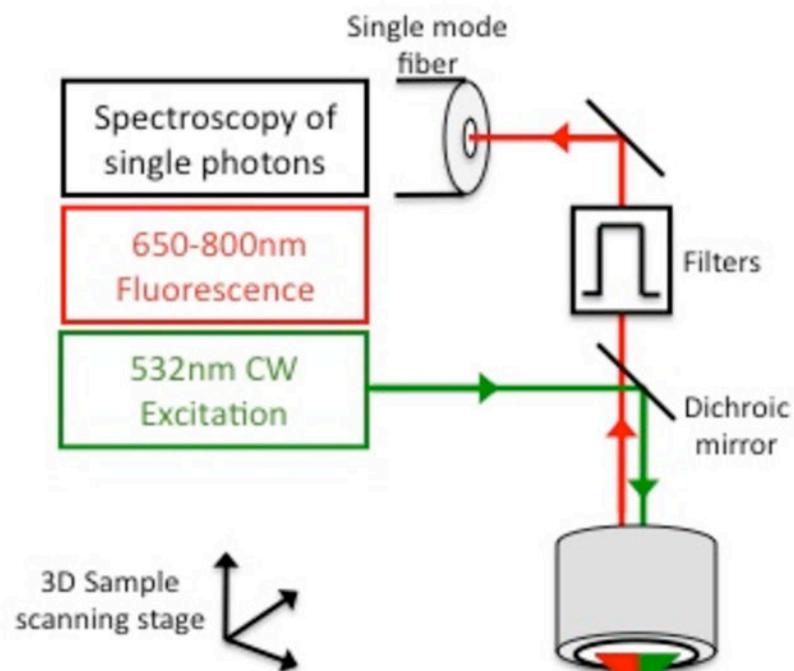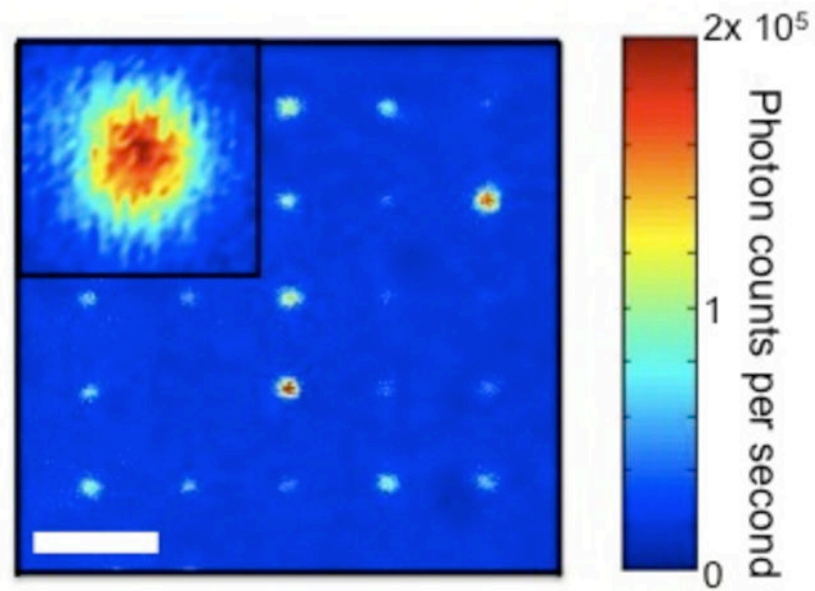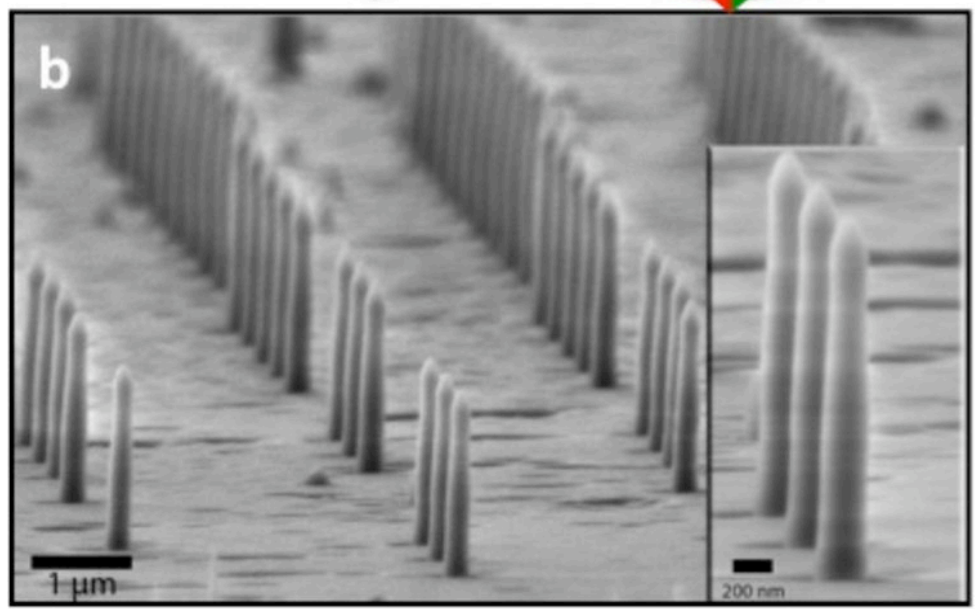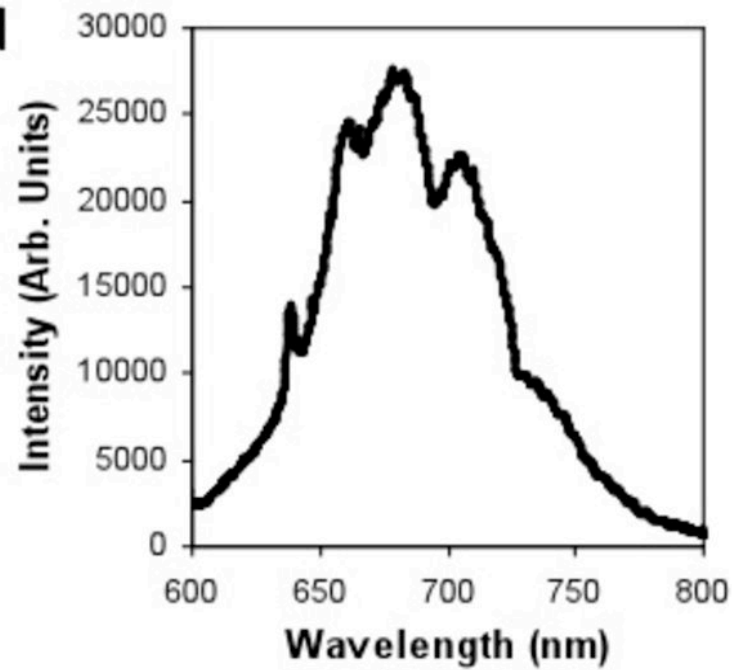

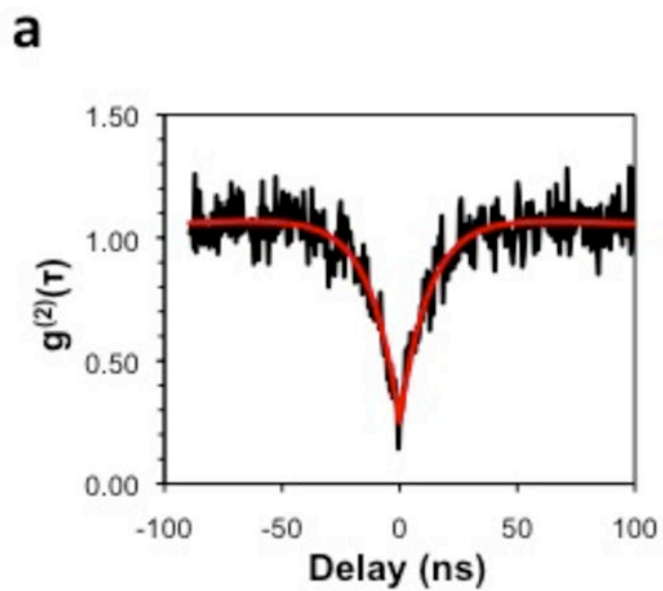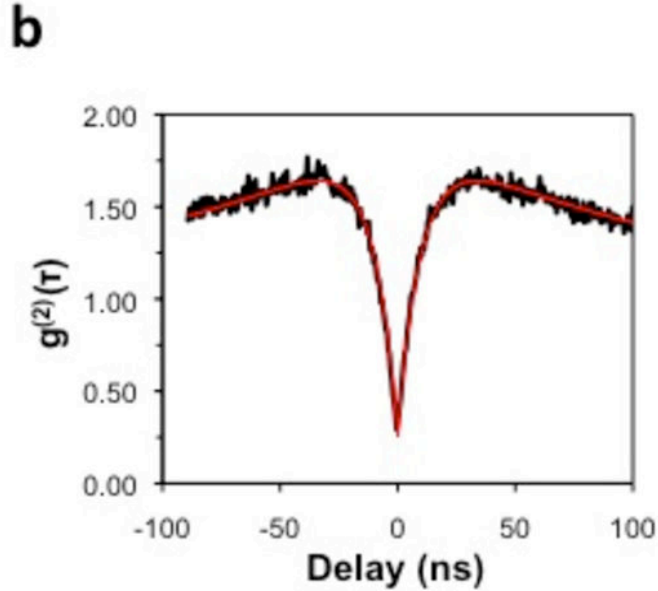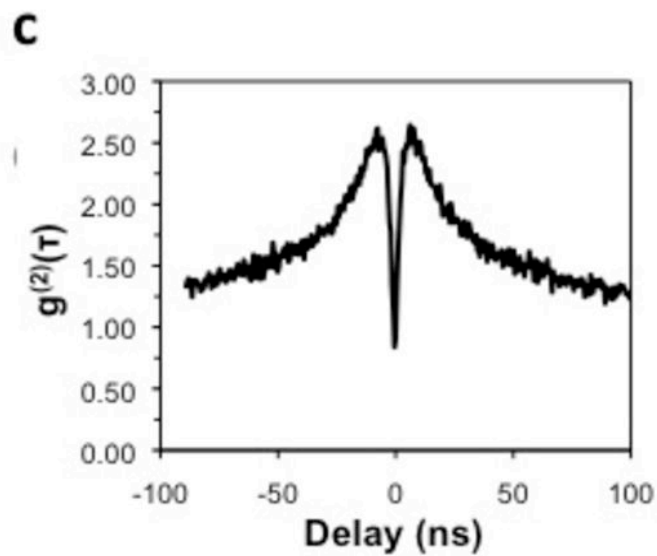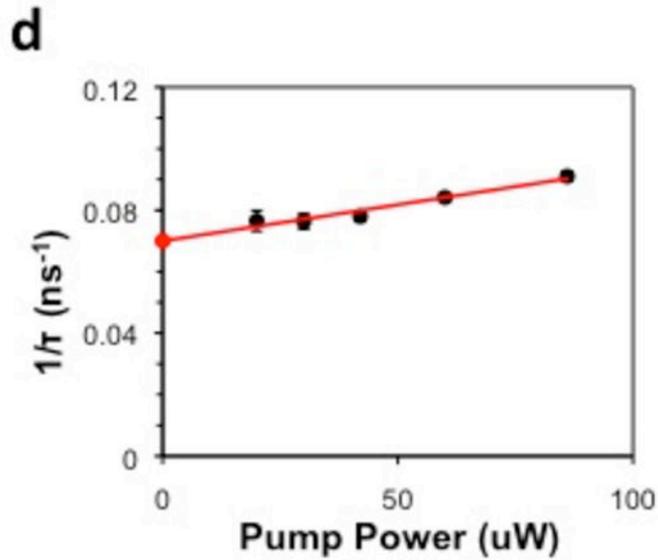

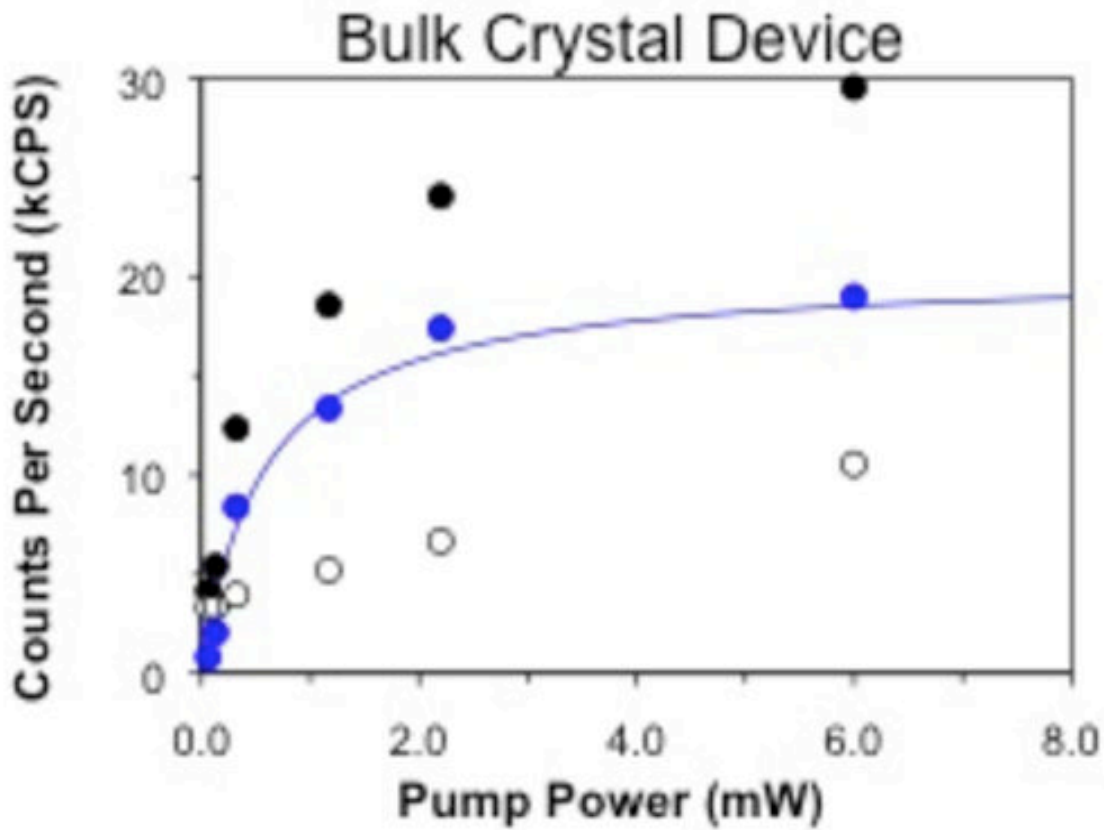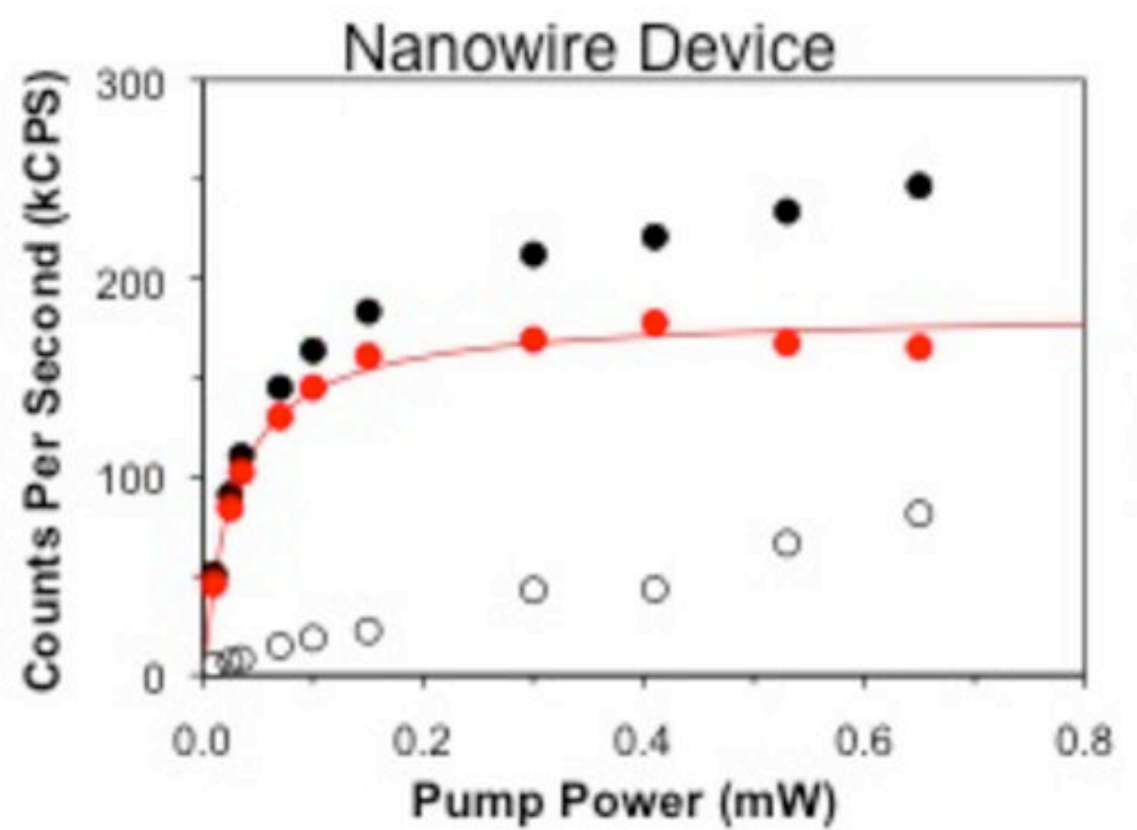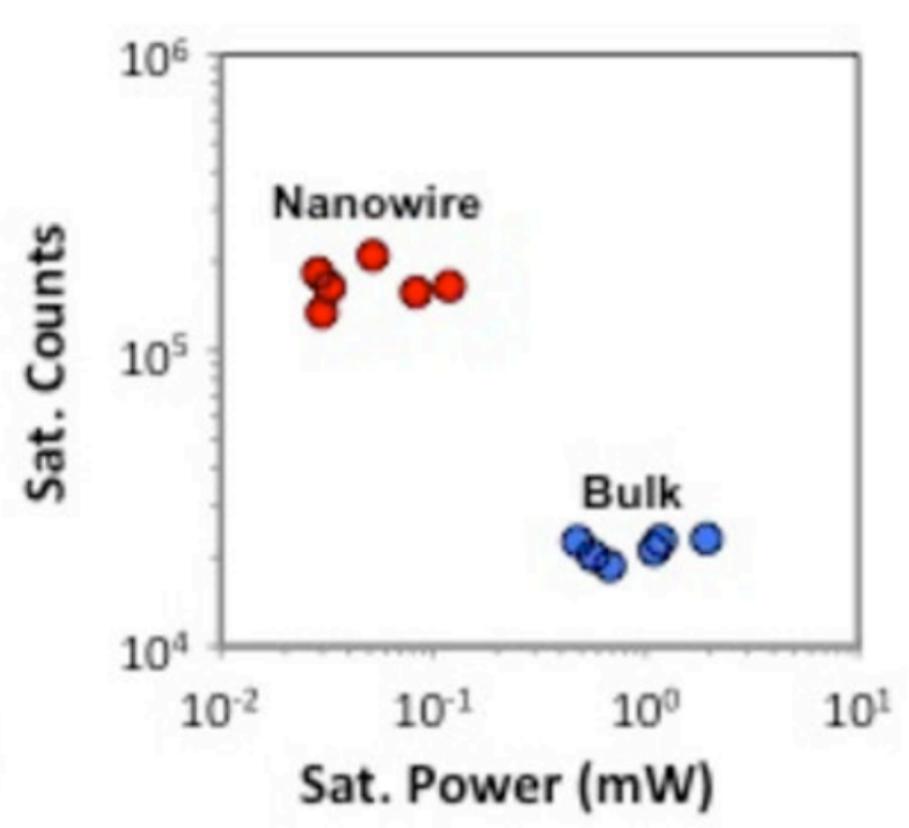